\documentclass{appolb}
\usepackage{epsfig}
\usepackage{subfigure}

\usepackage{graphicx}
\usepackage{amsmath}

\begin{document}
\title{Medium effects on meson couplings \\
 and dilepton spectra \thanks{Supported by the Polish State Committee for Science Research, grant 2 P03B 13324.}
\thanks{Talk presented at the 3rd Budapest Winter School on Heavy Ion Physics, 8-11 December 
2003, Budapest, Hungary.}}
\author{Agnieszka Bieniek 
\address{The H. Niewodnicza{\'n}ski Institute of Nuclear Physics Polish Academy of Sciences, 
ul. Radzikowskiego 152 PL-31342 Cracow, Poland}}
\maketitle

\begin{abstract}
{
The $\pi \omega \rho$ coupling constant is calculated in a relativistic hadronic framework 
incorporating nucleons and $\Delta(1232)$ isobars. Medium modifications are included. 
The vertex is analyzed in context of the
$\omega \to \pi^0  e^{+}e^{-}$ and $\rho^{a} \to \pi^{a} e^{+}e^{-}$ Dalitz 
decays in nuclear matter. A sizeable increase of the widths for these decays 
is found for invariant mass of dileptons in the range $0.2-0.5~{\rm GeV}$. This increase may help, 
among other medium effects, to explain the problem of the low-mass dilepton 
enhancement seen in relativistic heavy-ion collisions.} 
\end{abstract}
\PACS{25.75.Dw; 21.65.+f; 14.40.-n}

\vspace{0.1cm}

One of the major puzzles in relativistic heavy-ion experiments is 
the low-mass dilepton enhancement \cite{cereshelios}. This enhancement is observed in central nucleus-nucleus 
collisions (Pb-Au, S-Au) at high energies and is not present in proton induced reactions (p-Be, p-Au). 
The results of the CERES collaboration, taken from \cite{PhD}, are shown in Fig. 1. A visible increase 
of dilepton production is observed in the region between $0.2-0.6~{\rm GeV}$ of the invariant mass 
of dileptons. At energies of $158~{\rm A~GeV}$ or $200~{\rm A~GeV}$ 
the difference between the experiment and the cocktail model  
is of about a factor of 6. The cocktail model, designed to describe the p-p collisions, 
works very well for light-heavy systems,  but it fails to describe N-N collisions. 
The enhancement is also found 
at the lower energy of $40~{\rm A~GeV}$, where it is even more pronounced, see the right panel of Fig. 1.
\begin{figure}[htb]
\centerline{
\vspace{0mm} ~\hspace{0cm} 
\epsfxsize = 6 cm \centerline{\epsfbox{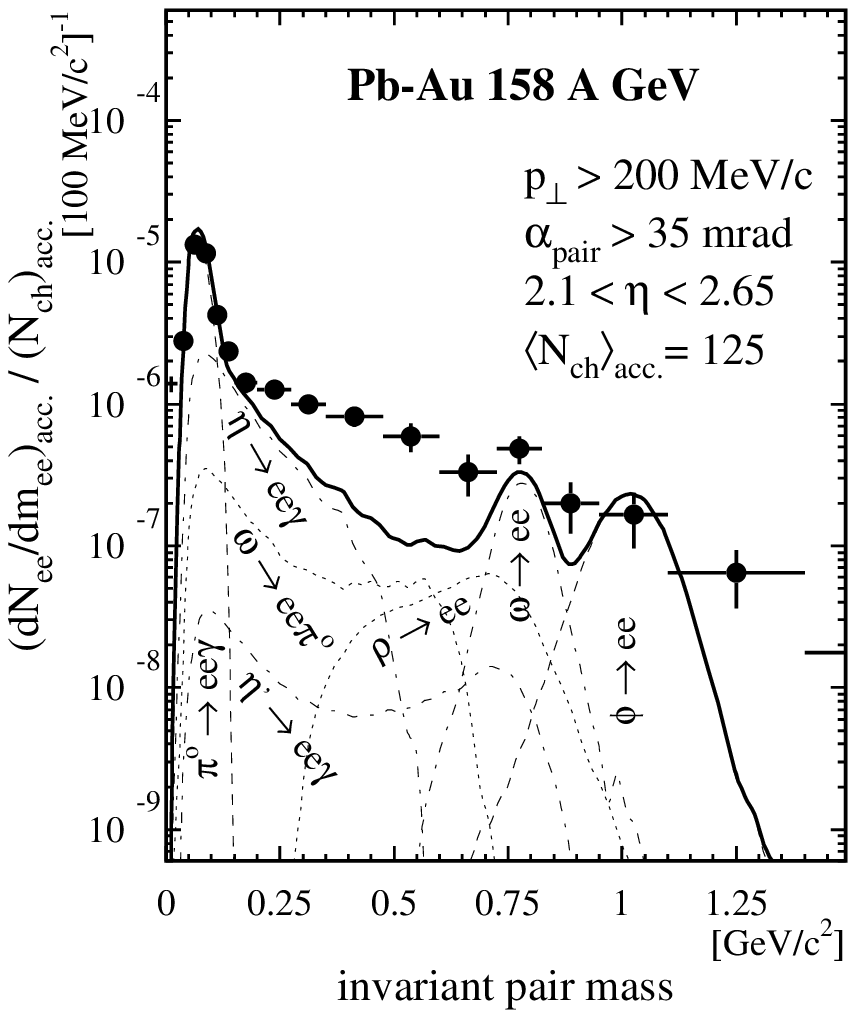}} \vspace{0mm}
\vspace{0mm} ~\hspace{-7.0cm}
\epsfysize = 6 cm \centerline{\epsfbox{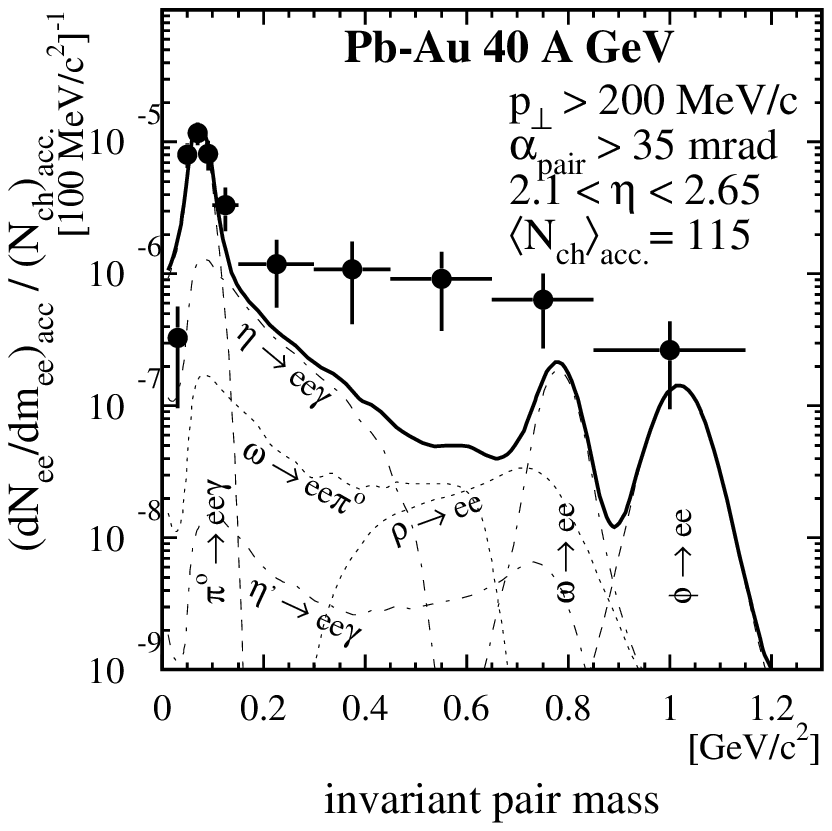}} \vspace{0mm}} 
\caption{Left: dilepton invariant mass spectrum of Pb-Au collisions at $158~{\rm A~GeV}$ 
from the CERES collaboration. The data is compared to the expected cocktail of hadronic sources. 
Right: the same at the lower energy of $40~{\rm A~GeV}$.}
\end{figure}

There are two major sources of dileptons: the direct decays of vector mesons, such as 
$\rho \rightarrow e ^{+}e ^{-}$, $\omega \rightarrow e ^{+}e ^{-}$, $\phi \rightarrow e ^{+}e ^{-}$, 
and the Dalitz decays, in particular $\omega \rightarrow \pi^{0} e ^{+}e ^{-}$, 
or $\rho^{a} \rightarrow  \pi^{a} e^{+}e^{-}$, which are the focus on this work.
Theorists have proposed possible explanations of the dilepton puzzle through 
in-medium modifications of hadron properties, in particular via lowering of the $\rho^{0}$ mass 
and/or broadening of the peak. By the way, next year we will get new data from NA60 experiment 
which will provide new insight into this problem \cite{bialk}.

Indications for in-medium physics come from RHIC \cite{fachini}, where the data obtained by the STAR 
collaboration have been interpreted as the lowering of the position of $\rho^{0}$ meson peak. 
Dynamical interactions with surrounding matter or the interference between different $\pi^{+}\pi^{-}$ 
scattering channels are possible explanations to the shift of the $\rho^{0}$ mass.

It has been well known to nuclear physicists that properties of hadrons such as mass, 
width, or size can be substantially altered in dense nuclear medium. 
The modification is due to direct interactions with the surrounding nucleons. 
Since the masses or widths of hadrons can be 
significantly modified we expect that coupling constants are also changed. For instance, it has been 
shown in \cite{wbwfbh, SongKoch, UBRW1, UBRW2,hiller} 
that the medium effects on the $\rho \pi \pi$ coupling constant are large. 
It has immediate consequences for the direct decays of the $\rho$ meson into dileptons. In this work 
we concentrate on the $\pi \omega \rho $ coupling constant and analyze its significance for the 
Dalitz decays of vector mesons. For more details see Ref. \cite{aa}.

\begin{figure}[htb]
\centerline{
\epsfysize = 3. cm \centerline{\epsfbox{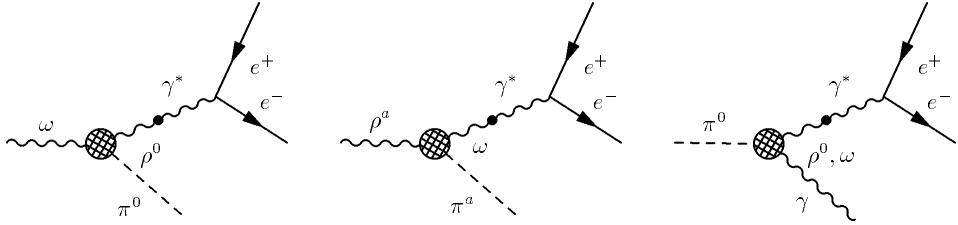}} \vspace{0mm}} 
\label{dalitz}
\caption{The processes where the $\pi \omega \rho$ vertex appears. Wavy lines indicates 
the $\rho$ or $\omega$, 
dashed lines the pion, the hatched blob denotes the medium-modified vertex, and 
the black dot is the vector-meson-dominance constant $\frac{em_{\omega }^{2}}{g_{\omega }}$ 
or $\frac{em_{\rho }^{2}}{g_{\rho }}$.}
\end{figure}
In Fig. 2, we show the $\pi \omega \rho $ vertex which enters the Dalitz decays 
$\omega \rightarrow \pi ^{0} e^{+}e^{-}$, $\rho \rightarrow \pi e^{+}e^{-}$, and 
$\pi ^{0}\rightarrow \gamma e^{+}e^{-}$. These processes are important for theoretical studies of the dilepton
production via Dalitz decays \cite{cereshelios}.
For instance the $\omega $ meson decays into $\pi ^{0}$ and $\rho ^{0}$, next $\rho
^{0}$ converts into a virtual photon, according to the vector
dominance principle, and finally the photon decays into the $e^{+} e^{-}$ pair. 
A similar situation occurs for the $\rho^{0}$ and $\pi^{0}$. Any modification of the vertex in Fig. 2 
will therefore directly affect the dilepton production rate.
\begin{figure}[htb]
\centerline{
\epsfysize = 3. cm \centerline{\epsfbox{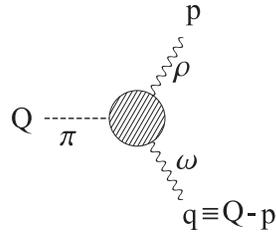}} \vspace{0mm}} 
\label{not}
\caption{The assignment of momenta in the $\pi \omega \rho$ vertex.}
\end{figure}

Through our analysis we use conventionally $Q$ as the incoming momentum of the pion, 
$p$ as the outgoing momentum of
the $\rho$, and $q = Q-p$ as the outgoing momentum of the $\omega $, see Fig. 3. 
With this convention the vacuum value of the $\pi \omega \rho$ vertex has the form
\begin{equation}
-iV_{\pi \omega^\mu \rho^\nu }=i\frac{g_{\pi \omega \rho }}{F_{\pi }}%
\epsilon ^{\mu \nu p Q},  \label{vac}
\end{equation}
where $\epsilon ^{\mu \nu pQ}=\epsilon ^{\mu \nu \alpha \beta }p_{\alpha }Q_{\beta }$ and
$F_\pi=93~{\rm MeV}$ denotes the pion decay constant. The vector mesons are, in general, virtual.
 
Our calculation of the in-medium $\pi \omega \rho$ vertex is made in the framework of 
a fully relativistic hadronic theory, where mesons interact with the nucleons and $\Delta$ isobars. 
The $\Delta$-resonance has a large influence on modification of the considered 
coupling because of a large coupling constant. 
For simplicity we work at the leading baryon density and at zero temperature. The vertex 
includes a vacuum part and a medium part. In the medium part there appear three types of diagrams 
with nucleons and $\Delta$-resonances, displayed in Fig. 4.
\begin{figure}[h]
\includegraphics[width=1.0\textwidth]{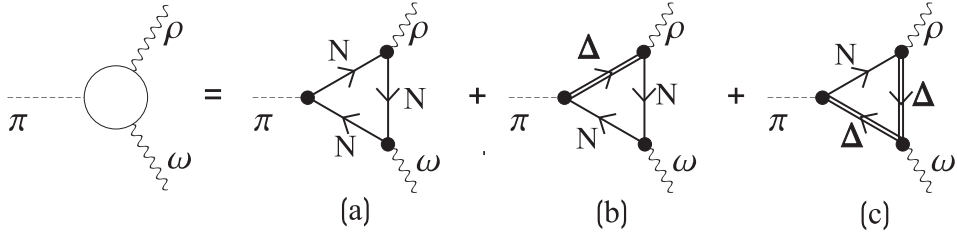}
\caption{\label{meddiag} In-medium diagrams included in present calculation 
(crossed diagrams not shown). Solid lines
indicates the in-medium nucleon propagator, double lines the $\Delta$ propagator.} 
\end{figure} 

The solid lines in Fig. 4 denote the nucleon propagator in nuclear matter, which is decomposed into the {\em %
free} and {\em density} parts in the usual way,
\begin{equation}
iS(k)  
= i(\gamma^{\mu} k_{\mu} + m_{N})[\frac{1}{k^{2} - m_{N}^{2} + i \epsilon }+ \\
\frac{i\pi }{E_{k}}\delta (k_{0}-E_{k})\theta (k_{F}-|k|)],  
\label{S}
\end{equation}
where $m_{N}$ denotes the nucleon mass, $E_{k}=\sqrt{m_{N}^{2}+k^{2}}$, and $%
k_{F}$ is the Fermi momentum. 
The Rarita-Schwinger propagator is used for the spin 3/2 $\Delta(1232)$ particle. 
\begin{equation}
iS_{\Delta }^{\mu \nu }(k)=i\frac{\gamma ^{\mu }k_{\mu }+M_{\Delta }}{%
k^{2}-M_{\Delta }^{2}}(-g^{\mu \nu }+\frac{1}{3}\gamma ^{\mu }\gamma ^{\nu }+%
\frac{2k^{\mu }k^{\nu }}{3M_{\Delta }^{2}}+\frac{\gamma ^{\mu }k^{\nu
}-\gamma ^{\nu }k^{\mu }}{3M_{\Delta }})  \label{rarita},
\end{equation}
we incorporate phenomenologically the effects of the non-zero width of the $
\Delta $ by the replacement $M_{\Delta }\rightarrow M_{\Delta }-i\Gamma /2$.
We use the standard form of the meson-baryon vertices, 
\begin{eqnarray}
-iV_{\omega ^{\mu }NN} &=& ig_{\omega }\gamma ^{\mu },\nonumber \\
-iV_{\rho _{b}^{\mu }NN} &=& i\frac{g_{\rho }}{2}(\gamma ^{\mu }- \frac{
i\kappa _{\rho }}{2m_{N}}\sigma ^{\mu \nu }p_{\nu })\tau ^{b}, \\
-iV_{\pi ^{a}NN} &=& \frac{g_{A}}{2F_{\pi }} Q_{\mu}\gamma^{\mu} \gamma ^{5}\tau ^{a}.
\nonumber 
\end{eqnarray}
For the vertices involving the $\Delta$ resonance we take
\begin{eqnarray}
-iV_{N\Delta ^{\alpha }\pi ^{a}} &=&g_{N\Delta \pi }Q^{\alpha }T^{a},  
\nonumber \\
-iV_{N\Delta ^{\alpha }\rho _{b}^{\mu }} &=&ig_{N\Delta \rho }(p_{\mu} \gamma^{\mu}
\gamma ^{5}g^{\alpha \mu }-\gamma ^{\mu }\gamma ^{5}p^{\alpha })T^{b}, \\
-iV_{\Delta ^{\alpha }\Delta ^{\beta }\omega ^{\mu }} &=&-ig_{\omega
}(\gamma ^{\alpha }\gamma ^{\mu }\gamma ^{\beta }-\gamma ^{\beta }g^{\alpha
\mu }-\gamma ^{\alpha }g^{\beta \mu }+\gamma ^{\mu }g^{\alpha \beta }), 
\nonumber
\end{eqnarray}
where $T^{a}$ are the standard isospin $\frac{1}{2}\rightarrow 
\frac{3}{2}$ transition matrices.
Below we give the tensor structure of the $\pi\omega\rho$ vertex in nuclear matter.
In the vacuum only the first structure appears, but in the medium, because of additional 
available four-vector, namely the four-velocity of the medium, $u$, 
we obtain nine structures in the amplitude,
\begin{eqnarray}
A^{\mu \nu } &=&A_{1}\varepsilon ^{\mu \nu pq}+A_{2}\varepsilon ^{\mu \nu
uq}+A_{3}\varepsilon ^{\mu \nu pu}+A_{4}\varepsilon ^{\mu upq}p^{\nu }
\label{struct} \\
&+&A_{5}\varepsilon ^{\mu upq}q^{\nu }+A_{6}\varepsilon ^{\mu upq}u^{\nu
}+A_{7}\varepsilon ^{\nu upq}p^{\mu }+A_{8}\varepsilon ^{\nu upq}q^{\mu
}+A_{9}\varepsilon ^{\nu upq}u^{\mu }.  \nonumber
\end{eqnarray} 
This structure follows from restrictions of Lorentz invariance and parity.

When the decaying particle is at rest with respect to the medium the four vector $q$ is parallel
to $u$. 
Because of that only two structures from Eq.~\ref{struct}, $A_{1}$ and $A_{3} \sim A_{1}$, are present. 
Combining the vacuum and medium pieces we have the following structure
\begin{eqnarray}
A^{\mu \nu }=\frac{i}{F_{\pi }}\frac{e^{2}}{%
g_{\rho }g_{\omega }}\left( g_{\pi \rho \omega }+\rho _{B}B\right) \epsilon
^{\nu \mu pq}, \label {amini}
\end{eqnarray}
which looks like a renormalization of the coupling constant. 
The quantity $B$ contains all medium contributions of diagrams from Fig. 4. 
Next, we introduce a convenient measure of the medium effects $g_{\rm eff}$ defined as
\begin{equation}
g_{\mathrm{eff}}=g_{\pi \rho \omega }+\rho _{B}B,  \label{geff}
\end{equation}
where $\rho_{B}$ is the baryon density. We investigate the dependence of $g_{\rm eff}$ on 
the dilepton invariant mass, $M_{e+e-}$. Our results are shown in Fig. 5, where   
\begin{figure}[htb]
\centerline{
\epsfysize = 9. cm \centerline{\epsfbox{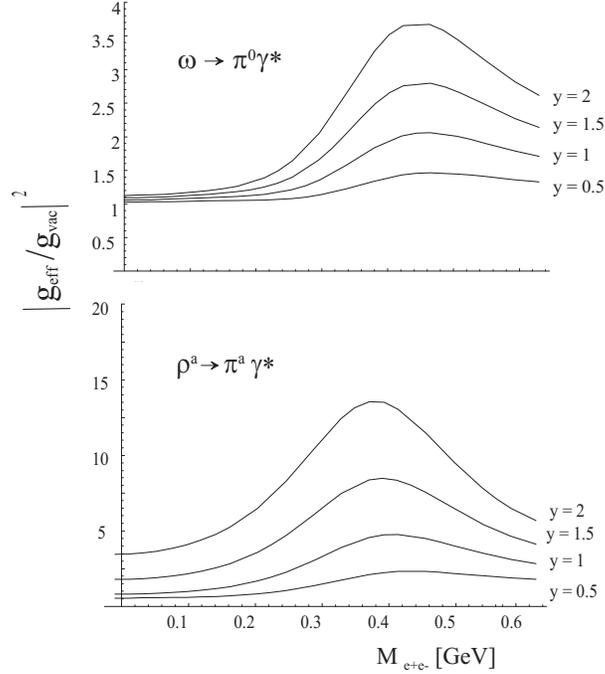}} \vspace{0mm}} 
\label{ly}
\caption{The $|g_{\rm eff}/g_{\rm vac}|^2$ for the $\protect%
\omega \to \protect\gamma^\ast \protect\pi^0$ and 
$\protect \rho^{a} \to \protect\gamma^\ast \protect\pi^a$ decays, 
for various value of the baryon density $y$, 
plotted as a function of the invariant mass of the dilepton, $M_{e+e-}$.}
\end{figure}
we plot $g_{\rm eff}^2$ 
divided by the vacuum value 
for various values of the baryon density $\rho_{B}=y \rho_{0}$, where $\rho_{0}=0.17~{\rm fm^{-3}}$ 
is the nuclear matter saturation density. It is visible that above $M_{e+e-}=0.2~{\rm GeV}$ for both 
$\omega \rightarrow \pi ^{0} e^{+}e^{-}$ and $\rho^{a} \rightarrow \pi^{a} e^{+}e^{-}$ decays 
the effective coupling constant is much larger than in the vacuum. 
Around $M_{e+e-}=0.4~{\rm GeV}$, at the saturation density ($y=1$), there is an enhancement 
of about a factor of 2 and 5 for the $\omega$ and $\rho$ meson, respectively. 
Below the value of $M_{e+e-}=0.2~{\rm GeV}$ the effective coupling constant is practically not changed. 
Obviously, the effect grows with $y$. 

Similar conclusions are obtained for the decaying particle moving with respect to the medium. In this case 
the properties of $\omega$ and $\rho$, in particular their widths, are different for transverse 
and longitudinal polarizations. Investigating the dependence of the width for both polarizations as a 
function of $M_{e+e-}$, we obtain that the medium effect weakens with growing 
momentum of the decaying particle. However, it remains substantial for 
momenta lower than $150~{\rm MeV}$, typical in a fireball formed in relativistic 
heavy-ion collisions. 
\begin{figure}[htb]
\centerline{
\vspace{0mm} ~\hspace{0cm} 
\epsfxsize = 7.2 cm \centerline{\epsfbox{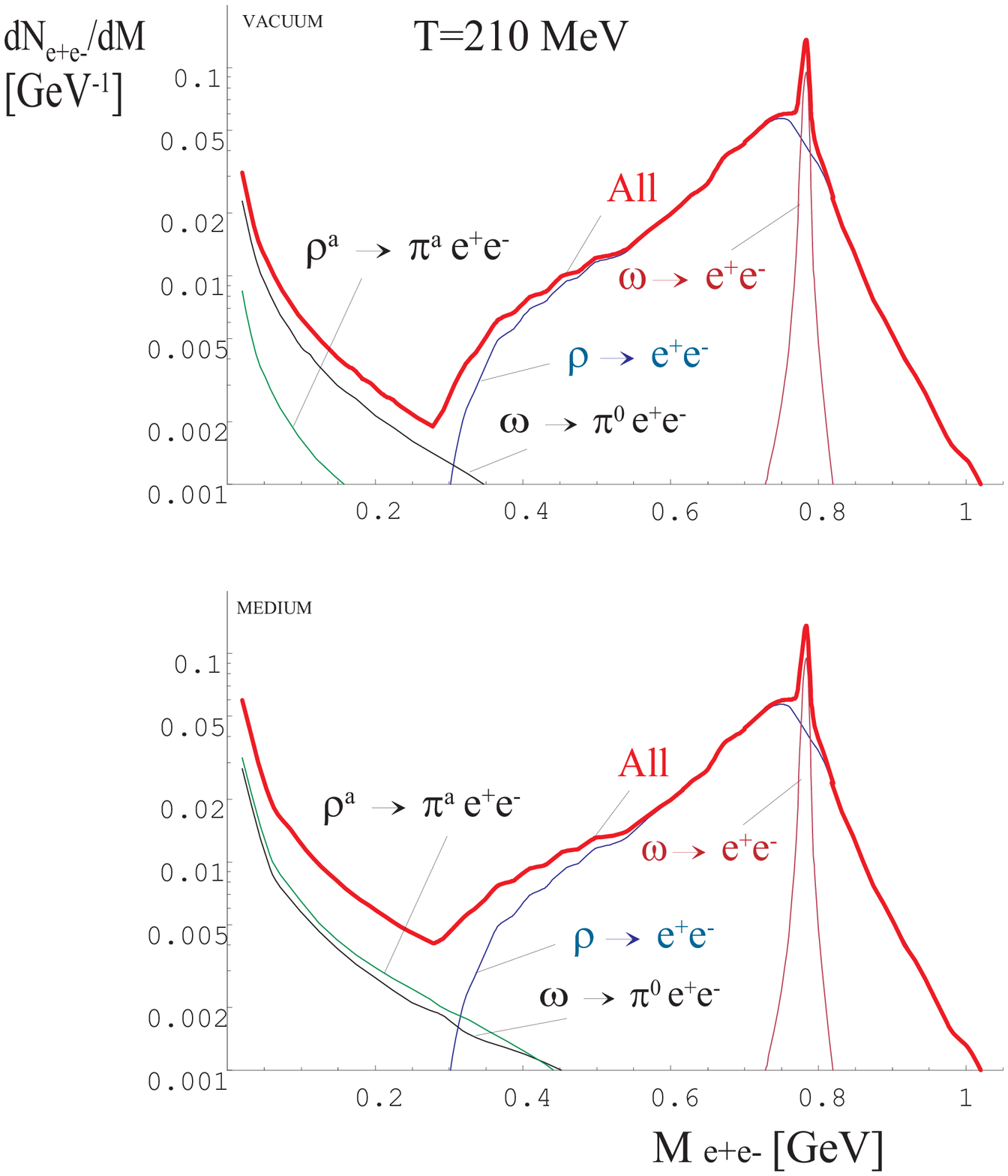}} \vspace{0mm}
\vspace{0mm} ~\hspace{-6.5cm}
\epsfysize = 8.55 cm \centerline{\epsfbox{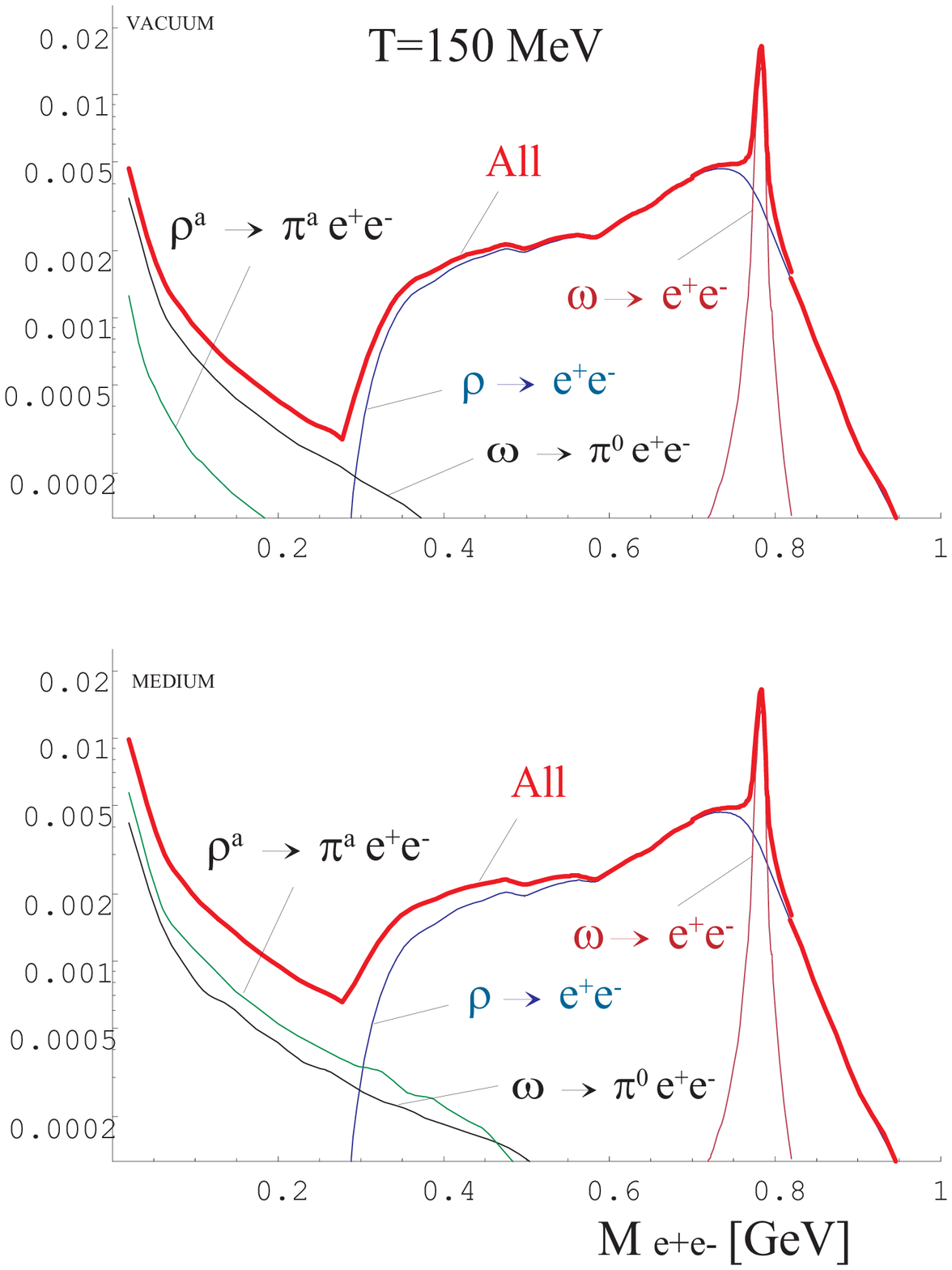}} \vspace{0mm}} 
\caption{The dilepton yield  for the direct decays as $\omega \to e^{+}e^{-}$ and 
$\rho \to e^{+}e^{-}$, and the Dalitz decays
$\omega \to  \pi^0 e^{+}e^{-}$ and $\rho^a \to \pi^a e^{+}e^{-}$, 
plotted as a function of the invariant mass of dileptons, $M_{e+e-}$. Two upper plots 
are done for the vacuum value of coupling constant, lower plots includes medium modified 
$\pi\omega\rho$ vertex. We show our results for two different temperatures,   
$T=210~{\rm MeV}$ (left) and $T=150~{\rm MeV}$ (right). 
The thick line indicates the sum of all contributions.}
\end{figure}

We have applied our model to evaluate the dilepton production from Dalitz decays given for the case 
without the Bose enhancement (neglected for brevity) \cite{PKoch} by 
\begin{equation}
\frac{dN_{e^{+}e^{-}}}{d^{4}x\,dM^{2}}=\int \frac{d^{3}p_{v}}{(2\pi
)^{3}E_{v}}\,f(p_{v })\frac{m_{v }}{\pi M^{3}}(2\Gamma
_{v \rightarrow \pi ^{0}\gamma ^{\ast }}^T+\Gamma
_{v \rightarrow \pi ^{0}\gamma ^{\ast }}^L)\,\Gamma _{\gamma ^{\ast
}\rightarrow e^{+}e^{-}}.  \label{PK}
\end{equation}
Here $v$ labels the vector meson, $f(p_v)=exp\left(\frac{-E_v}{T}\right)$ is the  
distribution function, and $\Gamma_{\gamma ^{\ast
}\rightarrow e^{+}e^{-}}$ is the partial decay width for the virtual photon. The 
transverse and longitudinal widths follow from our model. We see that an increased value of 
$\Gamma _{v \rightarrow \pi ^{0}\gamma ^{\ast }}$ results in an increased dilepton yield.
The lepton pairs are formed in the fire cylinder. We calculate the dilepton spectrum using
the model of
the hydrodynamic expansion of the fire cylinder from Ref. \cite{RappWamb},
which includes longitudinal and transverse expansion. Finally, the dilepton 
production rate is given by the formula 
\begin{eqnarray}
\frac{dN_{e+e-}}{dM} = \int\limits_0^{t_{max}} dt \int\limits_0^{r_{max} (t)} 2\pi r dr \int\limits_{-z_{max}
(t)}^{z_{max}(t)} dz \left( \frac {dN_{e+e-}}{d^4 x dM}\right),  
\end{eqnarray}
which is obtained from Eq.~\ref{PK} via integrating over the fireball.

Our numerical results 
are shown in Fig. 6. On the left-hand side we show the results for temperature 
$210~{\rm MeV}$ of Ref. \cite{RappWamb}. 
The upper plot is for the vacuum coupling constant. We display the direct contributions, such as 
$\omega \to e^{+} e^{-}$ and $\rho \to e^{+} e^{-}$, and the Dalitz decays of both $\omega$ and 
$\rho$ mesons. On the lower plot we present the same results but with medium modifications of the $\pi\omega\rho$ 
vertex. We see that the Dalitz decays are significantly enhanced. On the right-hand side we show 
the results for a lower temperature of $150~{\rm MeV}$, with similar conclusions. We stress that even 
in the vacuum the Dalitz decays are important in the region of $0.2-0.5~{\rm GeV}$, where the existing 
calculations have serious problems to explain the experimental data \cite{cereshelios}. 
The medium modifications enhance the effect.
We note that in the medium the $\rho$ decay becomes important to the $\omega$ decay, which is 
due to stronger medium enhancement and the isospin degeneracy.

In summary, the large increase of the coupling constant is observed for the 
$\omega \to \pi^0 e^{+}e^{-} $ and $\rho^a \to \pi^a e^{+}e^{-}$ in the range 
$M_{e+e-}=0.2-0.5~{\rm GeV}$. It may help to explain the long standing dilepton puzzle. 
We wish to point out that the Dalitz decays of the $\rho$ meson should not be neglected. 
In order to compare our results to experimental numbers, detector acceptance and efficiency  
should be included.

\end{document}